\documentclass[twocolumn,showpacs,preprintnumbers,amsmath,amssymb]{revtex4}

\usepackage{graphicx}
\usepackage{dcolumn}
\usepackage{bm}

\begin{document}

\title{NMR as a probe of the relaxation of the magnetization in magnetic molecules}

\author{P. Santini$^1$, S. Carretta$^1$, E. Liviotti$^1$, G. Amoretti$^1$, P. Carretta$^2$, M. Filibian$^2$,
A. Lascialfari$^2$ and E. Micotti$^2$}
\address{$^1$INFM and Dipartimento di Fisica, Universit\`a di
Parma, I-43100 Parma, Italy}
\address{$^2$INFM and Dipartimento di Fisica
A. Volta, Universit\`a di Pavia, I-27100 Pavia, Italy}

\date{January 12, 2005}
\begin{abstract}
We investigate the time autocorrelation of the molecular
magnetization $M(t)$ for three classes of magnetic molecules
(antiferromagnetic rings, grids and nanomagnets), in contact with
the phonon heat bath. For all three classes, we find that the
exponential decay of the fluctuations of $M(t)$, associated with
the irreversible exchange of energy with the heat bath, is
characterized  by a single characteristic time $\tau (T,B)$ for
not too high temperature $T$ and field $B$. This is reflected in a
nearly single-lorentzian shape of the spectral density of the
fluctuations.\\We show that such fluctuations are effectively
probed by NMR, and that our theory explains the recent
phenomenological observation by Baek et al. (PRB70, 134434) that
the Larmor-frequency dependence of $1/T_1$ data in a large number
of AFM rings fits to a single-lorentzian form.
\end{abstract}

\pacs{76.60.Es,76.60.-k,75.60.Jk}

\maketitle

A central aspect of the physics of magnetic molecules is the way
molecular observables, and particularly the magnetization $M$, are
affected by interactions of the spins with other degrees of
freedom. When the latter behave as a heat bath, they cause
decoherence of the time evolution of molecular observables,
usually leading to relaxation dynamics and exponential time-decay
of equilibrium fluctuations. The understanding and
characterization of relaxation mechanisms in crystals containing
magnetic molecules is not only important for evident fundamental
reasons, but also because relaxation and, more generally,
decoherence, constitute a major obstacle in the envisaged
technological applications of these molecules. In particular,
relaxation of $M$ through phonons sets an upper bound\cite{review}
to the temperature range in which nanomagnets could be used as
classical bits in a memory. In the possible application of some
molecules as elementary units (qubits) in quantum information
processing (QIP)\cite{loss}, energy dissipation from the qubits to
the nuclear-spins heat bath is the main concern regarding the
efficiency of quantum logical
gates and the performance of the computation.\\
In this work, we show that for three important classes of magnetic
molecules (antiferromagnetic (AFM) rings, grids and nanomagnets),
the exponential decay of the fluctuations of $M(t)$ due to
spin-phonon interactions is dominated by a {\it single}
characteristic time $\tau(T,B)$ over a wide range of temperatures
and applied fields. While this was already known for the long-time
decay of $M$ in nanomagnets, our results show that this applies to
completely different classes of molecules, and holds on much
shorter time-scales as well. Indeed, in the considered $(T,B)$
regimes this single characteristic frequency $\lambda_0 = 1/\tau$
dominates  the fluctuations spectrum to such a large extent that
even experimental probes whose intrinsic frequency is very
different from $\lambda_0$  will nevertheless detect mostly this
single spectral contribution. In particular, we show that NMR
measurements of the longitudinal nuclear relaxation rate $1/T_1$
can be used as a direct probe of the fluctuations of $M$, and that
experimental $1/T_1$ data are fully consistent with our picture.
For instance, we explain the recent experimental observation
\cite{Baek} that the Larmor-frequency dependence of $1/T_1$ data
in a large number of AFM rings fits to a lorentzian form, and we
elucidate the meaning and behavior of the characteristic time
$\tau (T,B)$ of this lorentzian. In addition, we collected new
data on the Mn3x3 grid which give further support to our picture.
At last, we
reanalyzed NMR data on the prototype nanomagnet Fe$_8$.\\
Each magnetic molecule can be described by the spin Hamiltonian
$$
H=\sum_{i>j=1,N}J_{ij}{\bf s}_{i}\cdot {\bf s}_{j} +
\sum_{i=1,N}\sum_{k,q}b_k^q(i)O_k^q({\bf s_i}) +
$$
\vspace{-0.5cm}
$$
\sum_{i>j=1,N}{\bf s}_{i}\cdot {\bf D}_{ij}\cdot {\bf s}_{j}-
g\mu_B B S_{z},\eqno(1)
$$
where ${\bf s}_{i}$ are spin operators of the $i^{th}$ magnetic
ion in the molecule. The first term is the isotropic Heisenberg
exchange interaction. The second term describes local
crystal-fields (CFs), with $O_k^q({\bf s_i})$ Stevens operator
equivalents for the $i$-th ion\cite{abragam} and $b_k^q(i)$ CF
parameters. The third term represents dipolar anisotropic
intra-cluster spin-spin interactions. The last term is the Zeeman
coupling with an external field ${\bf B}$ with $S_{z}=\sum_i
s_{i,z}$. The Heisenberg contribution is usually largely dominant,
and therefore the energy spectrum of $H$ consists of a series of
level multiplets with an almost definite value of the total spin
$S$.\\ Much of the physics of nanomagnets such as Fe$_8$ can be
described to a good degree of approximation by retaining the
lowest $S$-multiplet (e.g., $S=10$ in Fe$_8$)\cite{fe8}. In AFM
rings such as Fe$_6$ and Cr$_8$ and in grids, several distinct
$S$-multiplets must be included for a good description. The ground
and first-excited multiplets have $S=0$ and $S=1$ in AFM
rings\cite{rings}, while they have $S=5/2$ and $S=7/2$ in the
Mn3x3 grid\cite{grids}. In rings and grids the anisotropic terms
in Eq. (1) have no qualitative effects for most observables,
(including those on which we focus here) and in most situations
(i.e. away from intermultiplet (anti)-crossings caused by ${\bf
B}$). In view of this, and since numerical calculations based on
the full Hamiltonian Eq. (1) are cumbersome, in these two classes
of systems we will use Eq. (1) neglecting the anisotropic terms.
In addition, large-scale numerical techniques must be used for
Mn3x3, resulting in a limitation in the range of $T$ which can be studied.\\
Our first goal is the calculation of the damping of equilibrium
fluctuations of the molecule magnetization $M(t)\equiv g S_z$ due
to interactions of the electron spins with a heat bath. We adopt
the well-established framework presented in \cite{blum} for the
irreversible evolution of the density matrix $\rho_{s t}(t)$ (in
the representation of the {\it exact} eigenstates of (1)) in
presence of a heat bath. Here we focus on time scales $\tau$ much
longer than the typical periods of free evolution of the system
($2\pi\hbar/(E_s-E_t)$, with $E_s$ and $E_t$ eigenvalues of (1)).
In this case, the so-called ``secular approximation" enables the
time evolution of the off-diagonal matrix elements $\rho_{s\neq
t}(t)$ to be decoupled from that of the diagonal elements $\rho_{s
s}(t) := p_s(t)$. In a frequency-domain picture (e.g., the outcome
of a scattering experiment), this corresponds to a clear
separation of inelastic (IE) and quasielastic (QE) spectral
contributions. Then the (single-lorentzian) broadening of IE
transitions merely reflects the lifetimes of the involved pair of
levels (basically determined by Fermi's golden rule applied to the
spin-bath interaction). These IE off-diagonal contributions to the
evolution equations for $M(t)$ will be neglected since we are here
interested in the subtler QE components of the dynamics. Only the
latter are detected by low-frequency probes such as
AC-susceptibility or NMR. \\
In the secular approximation the populations of the molecular
eigenstates, $p_s(t)$, evolve through the well-known rate (master)
equations
$$
\dot{p}_s(t) = \sum_t W_{s t}p_{t} (t), \eqno(2)
$$
where $W_{s t}$ is the $s t$ element of the rate matrix
$\mathbb{W}$, i.e. the probability per unit time that a transition
between levels $\vert t \rangle$ and $\vert s \rangle$ is induced
by the interaction with the heat bath ($W_{s s} =-\sum_t W_{t
s}$).\\
We are here focusing on temperatures for which relaxation is
determined by interactions of the spin degrees of freedom of each
molecule with the crystal phonon heat bath. Since we are
interested in general qualitative features of the relaxation
dynamics and not in compound-specific details, we will eventually
assume a unique simple form for the spin-phonon coupling for all
compounds. The main contribution to this coupling comes from the
modulation of the local rank-2 crystal fields
$\sum_{l}\sum_{q}b_2^q(l)O_2^q({\bf s}_l)$ by phonons :
$$
V=\sum_{i=l,N}\sum_{\Gamma_l ,\gamma_l} V_{\Gamma_l} Q_{\Gamma_l
,\gamma_l} C_{\Gamma_l ,\gamma_l} , \eqno(3)
$$
where $Q_{\Gamma_l ,\gamma_l}$ are symmetrized displacements of
neighboring atoms of the $l$th magnetic ion, labelled by
irreducible representations $\Gamma_l$ of the local point group of
ion $l$, $\gamma_l$ being the component. $C_{\Gamma_l ,\gamma_l}$
are linear combinations of $O_2^q({\bf s}_l)$ transforming as
${\Gamma_l ,\gamma_l}$. For instance, in octahedral local
environment, if $\Gamma \equiv \Gamma_3$, $C_{\Gamma ,1} =
O_2^0({\bf s}_l)/\sqrt{3}$ and $C_{\Gamma ,2} = O_2^2({\bf s}_l)$.
The symmetrized displacements $Q_{\Gamma_l ,\gamma_l}$ of the
neighboring atoms can be expanded in terms of normal coordinates
$q_{{\bf k}\sigma}$ of the full crystal lattice,
$$
Q_{\Gamma_l ,\gamma_l} = \sum_{{\bf k}\sigma} a_{{\bf k}\sigma}
(l) q_{{\bf k}\sigma}, \eqno(4)
$$
where $a_{{\bf k}\sigma} (l)$ are known as Van Vleck coefficients.
This leads to the following general form for the spin-phonon
coupling potential
$$
V=\sum_{l=1,N}\sum_{Q=-2,2}\sum_{{\bf k}\sigma} C_{Q}(l,{\bf
k},\sigma) O_2^Q({\bf s}_l) ( c_{{\bf k}\sigma} + c_{{\bf
-k}\sigma}^{\dagger}) ,\eqno(5)
$$
where $C_{Q}(l,{\bf k},\sigma)$ is the coupling constant between
the $Q$-type electric quadrupole on ion $l$ ($O_2^Q({\bf s}_l)$),
and phonon modes of wavevector ${\bf k}$ and branch index
$\sigma$\cite{gehring}. Clearly, a detailed microscopic
calculation of $C_{Q}(l,{\bf k},\sigma)$ is unfeasible: it would
require to model the (complicated) phonon spectrum, and to fix the
value of a great number of parameters $V_{\Gamma_i}$. Yet,
experimental information is completely insufficient to accomplish
this. Therefore, we have chosen as usual\cite{villain} to adopt a
Debye model for phonons, and for the sake of simplicity we assume
$Q=0,1,2$ with $C_{Q}(l,{\bf k})$ independent of $l$ and $Q$,
although compound-specific $C_{Q}(l,{\bf k},\sigma)$ would be
needed for precise quantitative calculations, which are beyond the
aim of this work. We have checked that the results do not change
qualitatively if different choices are made\cite{nota1}. \\The
rate matrix $\mathbb{W}$ appearing in Eq. (2) can be calculated by
first-order perturbation theory applied to $V$, leading to
$$
W_{s t}=\gamma \pi \vert \langle t \vert
\sum_{i=1,N}\sum_{Q=0,1,2} O_2^Q({\bf s}_i) \vert s \rangle \vert
^2 \Delta^3 _{s t} n(\Delta_{s t}) ,\eqno(6)
$$
with $ n(x)=(e^{\beta \hbar x}-1)^{-1},
\Delta_{st}=(E_s-E_t)/\hbar.$ In our simplified spin-phonon model
$\gamma$ is the unique free parameter and describes the
spin-phonon coupling strength. It will be determined by comparison
with experimental data.
\begin{figure}
\centering
\includegraphics[width=0.5\textwidth]{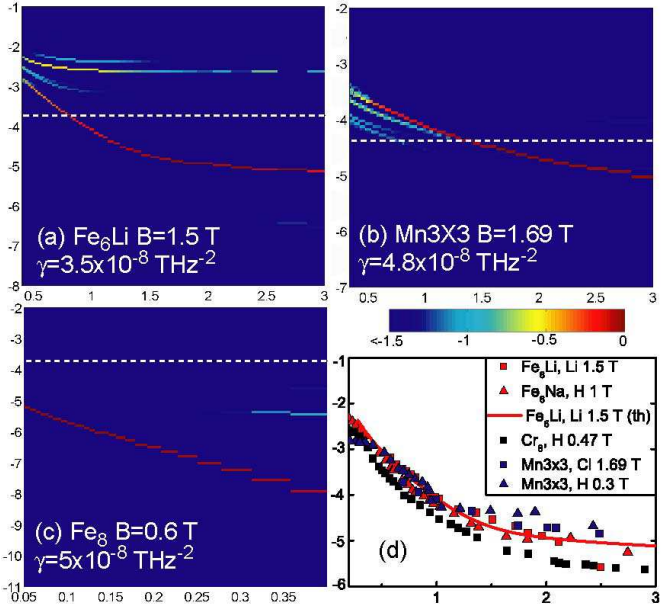}
\caption{\label{fig1} Calculated frequency weights
$A(\lambda_i,T,B)$ of the magnetization autocorrelation {\it vs}
$\Delta/T$ (a,b), with $\Delta$ the exchange gap, or {\it vs}
$1/T$ (c) ($x$-axis). The $y$-axis is $\log_{10}(\lambda )$ (in
THz). The color maps $\log_{10} (A(\lambda_i, T,B)/(\chi T))$ (see
Eq. (9)). The hamiltonian parameters are given in
\cite{fe8,rings,grids}. For each value of $T$, the spectra have
been normalized by $\chi T$, which is proportional to the size of
equilibrium fluctuations. The white dashed lines are the $^7$Li
(a), $^{35}$Cl (b), and $^1$H (c) Larmor angular frequencies. The
results for Cr$_8$ (not shown)
are similar to those of Fe$_6$Li, with $\gamma = 5\times 10^{-7}$Thz$^{-2}$.\\
Panel (d) shows the characteristic frequency $\lambda_0\; vs \;
\Delta/T$ for Cr$_8$, Fe$_6$Li, Fe$_6$Na and Mn3x3 extracted from
the experimental $1/T_1$ data (see text). The data are taken from
ref. \cite{Baek}, apart from those for Mn3x3 which are new. The
red line represents the dominant frequency $\lambda_0$ of Fe$_6$Li
deduced from panel (a).}
\end{figure}
The next step is the calculation of the Fourier-transform
$S_{\mathcal{A},\mathcal{B}}(\omega)$ of the cross-correlation
function $\langle \Delta \mathcal{A}(t)\Delta \mathcal{B}(0)
\rangle$ implied by Eq. (2) for equilibrium fluctuations of two
generic observables $\mathcal{A}$, $\mathcal{B}$. We exploit the
fluctuation-dissipation theorem, which implies
$$
S_{\mathcal{A},\mathcal{B}}(\omega) = \frac{2 k_{\mathrm B}T
\chi_{\mathcal{A},\mathcal{B}}^{\prime\prime}(\omega)}{\omega}=2
k_{\mathrm B}T
\;\tilde{R}^{\prime}_{\mathcal{A},\mathcal{B}}(i\omega),\eqno(7)
$$
with $\chi_{\mathcal{A},\mathcal{B}}^{\prime\prime}(\omega)$ the
imaginary part of the susceptibility, and
$\tilde{R}^{\prime}_{\mathcal{A},\mathcal{B}}(z)$ the real part of
the Laplace transform of the relaxation function
$R_{\mathcal{A},\mathcal{B}}(t)$. The latter is defined as
$\lim_{\epsilon\rightarrow 0}(\langle
\mathcal{A}(t)\rangle-\langle \mathcal{A}\rangle^{eq})/\epsilon$,
with a perturbation $\epsilon\mathcal{B}$ abruptly switched-off at
$t=0$. We have calculated $R_{\mathcal{A},\mathcal{B}}(t)$ by
solving the system of equations (2), obtaining the following
expression:
$$
S_{\mathcal{A},\mathcal{B}}(\omega) = \sum_{q,t}(\mathcal{B}_{tt}-
\langle \mathcal{B} \rangle ^{(eq)})
$$
\vspace{-0.7cm}
$$
\times(\mathcal{A}_{qq} - \langle \mathcal{A} \rangle
^{(eq)})Re\left( p_t^{(eq)} (\frac{1}{i\omega -
\mathbb{W}})_{qt}\right),  \eqno(8)
$$
where $p_t^{(eq)}$ is the equilibrium population of the $t$-th
level, and $\mathcal{B}_{tt} = \langle t \vert \mathcal{B} \vert t
\rangle$. This leads to correlations decaying as a sum of
exponentials with $n$ characteristic times $\tau_{QE}^{(i)}$ given
by the inverse eigenvalues of $-\mathbb{W}$, $n$ being the
dimension of the molecule spin Hilbert space. Therefore, the QE
frequency-spectrum is in general a sum of lorentzians centered at
zero frequency having width $1/\tau_{QE}^{(i)}$. It should be
stressed that the $n$ $\tau_{QE}^{(i)}$ cannot be directly
determined from the $n$ level lifetimes $\tau_{life}^{(s)}$.
Although both $\tau_{QE}^{(i)}$ and $\tau_{life}^{(s)}$ contain
integrated information on the spin-bath dissipation channels, this
information is different for the two sets. Thus, low-frequency
measurements probing the QE dynamics provide different and
complementary information with respect to direct lifetime
measurements by high-frequency
techniques (e.g., inelastic neutron scattering).\\
We focus now on the important case $\mathcal{A} = \mathcal{B} =
S_z$, i.e. we study the spectrum of fluctuations of the cluster
magnetization. In general, many different $1/\tau_{QE}^{(i)}$ may
be expected to contribute to this spectrum, i.e.
$$
S_{S_z ,S_z}(\omega ,T,B) = \sum_{i=1,n} A(\lambda_i,T,B)
\frac{\lambda_i(T,B)}{\lambda_i(T,B)^2 + \omega^2}, \eqno(9)
$$
where $\lambda_i(T,B)\equiv 1/\tau_{QE}^{(i)}$. Figure 1 shows
intensity plots of the frequency weights $A(\lambda_i,T,B)$ of the
magnetization autocorrelation, calculated from Eq. (8),  as
function of temperature $T$ for three representative systems. The
single free parameter $\gamma$ has been estimated by the observed
peak in the $T$-dependence of the NMR $1/T_1$ (see below), apart
from Fe$_8$ for which relaxation data allow to fix $\gamma$
directly. The precise value of $\gamma$ does not change
the spectra appreciably, but merely sets the frequency scale.\\
\begin{figure}
\centering
\includegraphics[width=0.5\textwidth]{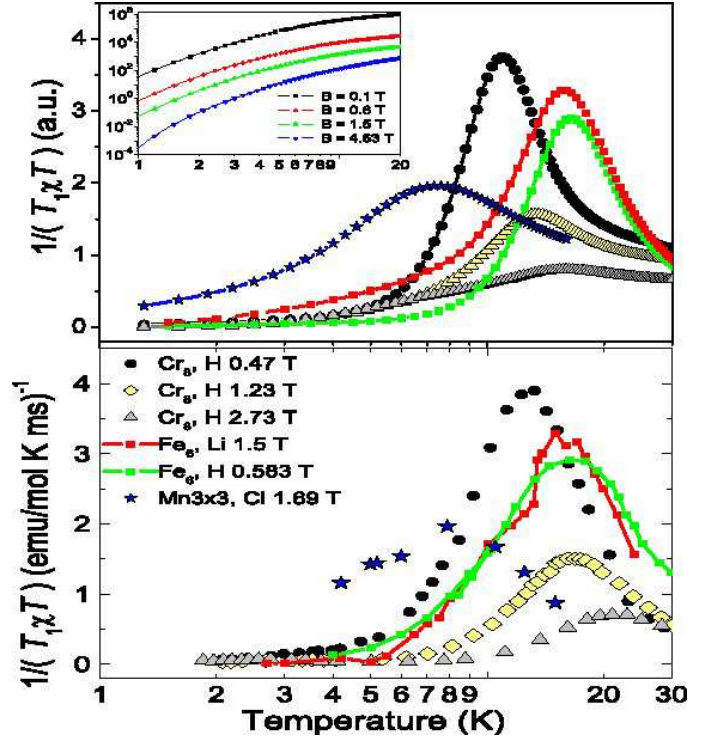}
\caption{\label{fig2} Up : calculated $1/T_1/(\chi T)$ as a
function of the temperature $T$ for Cr$_8$, Fe$_6$Li and Mn3x3 for
different fields and nuclei. Inset :  Calculated $1/T_1(T)$ (in
a.u.) for Fe$_8$ for different values of $B$ applied along the
easy axis. Down: Measured $1/T_1/(\chi T)$ {\it vs} $T$ for
Cr$_8$, Fe$_6$Li and Mn3x3. In order to fit the figure, data for
Fe$_6$Li were divided by 3, and those for Mn3x3 by 5.}
\end{figure}
The calculations reported in Fig. 1 show that for a fairly wide
range of values of $T$ and $B$, {\it a single frequency}
$\lambda_0(T,B)$ (and therefore a single lorentzian) dominates the
spectrum in the three classes of molecules we considered (AFM
rings, AFM grids and nanomagnets). The dominating frequency
$\lambda_0$ is different from the lifetime broadening of any
level. In the special case of nanomagnets (Fe$_8$, Mn$_{12}$), the
calculated $\lambda_0$ coincides with the smallest $\lambda_i$.
Hence, in this case $\lambda_0^{-1}$ is also the time determining
the asymptotic return to equilibrium of the magnetization when the
nanomagnet is prepared in a highly nonequilibrium state by a
strong applied field. Instead, our calculations show that in rings
and grids $\lambda_0$ {\it is orders of magnitude larger} than the
smallest $\lambda_i$ (the latter is not visible in Fig. 1 since
the associated weight is tiny). AFM rings are characterized by a
single dominating characteristic frequency $\lambda_0$ for $B\ll
B_{c}$ with $B_c$ the field at which an $S=1$ level crosses $S=0$,
and for $T\lesssim 12 J/N$. This corresponds to the regime in
which the magnetization and its fluctuations are mainly determined
by levels belonging to the $S=1$ manifold. In this
single-frequency regime, the decay of fluctuations takes places
mostly through indirect Orbach-like processes involving
higher-lyings $S$-manifolds, resulting in an approximately
exponential dependence in $1/T$ of $\lambda_0$ (see Fig. 1). A
different $T$-dependence occurs at low $T$, where direct
transitions within the field-split $S=1$ manifold constitute the
main relaxation mechanism. Fe$_8$ is also characterized by a
single dominating characteristic frequency $\lambda_0$, depending
exponentially on $1/T$, if the Zeeman energy is small with respect
to the anisotropy barrier. Finally, the Mn3x3 grid is again
characterized by a dominating frequency for $B\ll B_{c}$ with
$B_c$ the field at which an $S=7/2$ level crosses $S=5/2$, and for
$T\lesssim \Delta/k_{B}$, with $\Delta \sim 0.9$ meV the exchange
gap. Therefore, for all these systems the calculated dynamical
structure factor of Eq. (9) is dominated by a single contribution:
$$
S_{S_z,S_z}(\omega,T,B) \simeq A(\lambda_0,T,B)
\frac{\lambda_0(T,B)}{\lambda_0(T,B)^2 + \omega^2}.\eqno(10)
$$
From linear-response theory $A(\lambda_0,T,B)\propto \chi(B,T) T$,
with $ \chi(B,T)$ the  static Curie susceptibility. Therefore, a
measurement of $S_{S_z,S_z}$ allows the $T$- and $B$-dependence of
$\lambda_0(T,B)$ to be extracted straightforwardly. In the
following we will show that NMR confirms this picture and allows
$\lambda_0(T,B)$ to be
determined. \\
The nuclear spin-lattice relaxation rate $1/T_1$ can be calculated
by a perturbative treatment of the hyperfine interactions between
nuclear and electronic spins\cite{Moriya}. Within this framework,
$1/T_1$ is given by a linear combination of the Fourier transforms
of electronic spin correlation functions
$S_{s_{i,\alpha},s_{j,\beta}}$, evaluated at the Larmor frequency
$\omega_L(B)$. By considering only the Heisenberg and Zeeman terms
in $H$ (Eq. (1)), the spin lattice relaxation rate of a magnetic
nucleus of Larmor frequency $\omega_L$ interacting with a cluster
composed of N spins may be written as
$$
1/T_1 \propto  \sum_{i,j=1,N}  \alpha_{ij}
(S_{s_i^z,s_j^z}(\omega_L)
$$
\vspace{-0.7cm}
$$
+S_{s_i^z,s_j^z}(-\omega_L))+ \beta_{ij}
(S_{s_i^+,s_j^-}(\omega_L)+S_{s_i^+,s_j^-}(-\omega_L)
$$
\vspace{-0.7cm}
$$
+S_{s_i^-,s_j^+}(\omega_L) +S_{s_i^-,s_j^+}(-\omega_L)), \eqno(11)
$$
where $\alpha_{ij}$ and $\beta_{ij}$ are geometrical coefficients
of the dipolar interaction between nuclear and electronic spins.
Since generally $\omega_L \ll \Delta_{s t}$, only the QE part of
the $S_{A,B}(\omega )$ in the above formula contributes to
$1/T_1$. If anisotropy is neglected, Eq. (8) shows that only the
$\alpha = \beta = z$ terms are nonzero because $\langle t \vert
s_{i,x} \vert t \rangle = \langle t \vert s_{i,y} \vert t \rangle
= 0$. In addition, for a magnetic ring $\langle t \vert s_i^z
\vert t \rangle = \langle t \vert s_j^z \vert t \rangle = 1/N
\langle t \vert S_z \vert t \rangle$, where $S_z$ is the $z$
component of the {\it total} spin operator $\bf{S}$. Then,
$$
1/T_1 =  G \; S_{S_z,S_z}(\omega_L), \eqno(12)
$$
where $G$ is a scale factor depending only on the positions and
spins of the probed nuclei. The same result holds also for grids
as long as the ground multiplet gives the main contribution (i.e.
at sufficiently low $T$). In fact, in this regime all
$S_{s_{i,z},s_{j,z}}(\omega)\propto S_{S_z,S_z}(\omega)$. We have
numerically checked that this holds approximately in the
single-lorentzian regime (see Fig. 1). If magnetic anisotropy
cannot be neglected, as in the case of Fe$_8$, the simple result
of Eq. (12) can be recovered if ${\bf B}$ is along the easy axis,
by exploiting the fact that within the ground $S$-multiplet
$S_{s_{i,z},s_{j,z}}(\omega)\propto S_{S_z,S_z}(\omega)$.
Therefore, in all these systems NMR measurements of $1/T_1$ can be
used as a probe of $S_{S_z,S_z}(\omega)$.\\
Eqs.(10) and (12) imply that, as a function of $T$, $1/T_1/(\chi
T)$ should display a single peak when $\lambda(T,B) =
\omega_L(B)$. Moreover, since $\lambda(T,B)$ depends little on $B$
for not too large $B$, $1/T_1/(\chi T)$ measured on different
nuclei and by setting the value of $B$ in order to keep
$\omega_L(B)$ fixed, should display a peak at approximately the
same $T$. This is clearly visible, for instance, in our
calculations of $1/T_1/(\chi T)$ for Fe$_6$Li on Li and H nuclei,
shown in Fig. 2, which contains our calculated $T$-dependence of
$1/T_1$. By increasing $B$ by a factor $p$ and keeping the nucleus
fixed, the peak value of $1/T_1/(\chi T)$ should decrease by about
$p$. This can be seen in Fig. 2 in the calculated curves for
Cr$_8$. This is what had been experimentally observed in
\cite{Baek}, where a phenomenological single-lorentzian expression
for the Larmor-frequency dependence of $1/T_1$ had been shown to
fit NMR data in several AFM rings very well. Part of the data are
included in the lower panel of Fig. 2 together with new data we
collected on the Mn3x3 grid. The agreement with our calculations
is very good. In addition, panel (d) in Fig. 1 contains the
$T$-dependence of the characteristic frequency $\lambda_0$
extracted from the lorentzian fit of the data\cite{Baek}. The
results are in good agreement with the behavior of the calculated
$\lambda_0$ in panels (a,b). Similar agreement is obtained for the
Cr$_8$ ring. We remark that in the $T$-range where the peak in
$1/T_1$ occurs the calculated $T$-dependence of $\lambda_0$ is
exponential in $1/T$ (relaxation through intermultiplet Orbach
processes), and in good agreement with experimental data. Hence,
the phenomenological power-law T-dependence assumed in
\cite{Baek}, although fitting the data over a limited temperature
range, is not confirmed by our
microscopic calculation.\\
In the case of Fe$_8$, the free parameter $\gamma$ has been fixed
in order to reproduce the $T$-dependence of the relaxation time of
the magnetization in the thermally-activated regime measured in
\cite{sangregorio}. There is therefore no free parameter in the
calculation. As can be seen in Fig. 1, the calculated $\lambda_0
(T)$ is much smaller than the $^1$H Larmor frequency. Indeed,
measured $1/T_1$ data do not show a peak\cite{fe8nmr}, and are in
good agreement with our parameter-free calculations reported in the inset of Fig. 2.\\
In conclusion, a general expression for the quasielastic part of
the dynamic structure factor in presence of a heat bath has been
derived, starting from the master equation applied to the full
microscopic hamiltonian. By specializing this expression to the
case of the total molecule magnetization, we have calculated (Fig.
1) the frequency spectral weights of the decay of fluctuations of
this observable. We have shown that in AFM rings, grids and
nanomagnets the decay of equilibrium fluctuations due to
interactions with phonons is characterized by a single
characteristic frequency $\lambda_0(T,B)$ over a wide range of
temperatures and fields. We have also shown that the NMR
spin-lattice relaxation rate $1/T_1$ can be used to measure
$\lambda_0(T,B)$.  This microscopic result differs from the
previous phenomenological models of $1/T_1$ in rings\cite{Baek}
and nanomagnets\cite{fe8nmr}, where $1/T_1$ was assumed to probe
level lifetimes in presence of phonons, and not a relaxation time.
Our calculations are in good agreement with existing experimental
NMR data, and with our new measurements on the Mn3x3 grid.\\
Useful discussions with F. Borsa are gratefully acknowledged. We
thank L.K. Thompson for the synthesis of Mn3x3, for which
preliminary NMR results are published here.

\end{document}